\documentclass[fleqn,usenatbib]{mnras}

\usepackage{newtxtext,newtxmath}
\usepackage[T1]{fontenc}

\DeclareRobustCommand{\VAN}[3]{#2}
\let\VANthebibliography\thebibliography
\def\thebibliography{\DeclareRobustCommand{\VAN}[3]{##3}\VANthebibliography}

\usepackage{graphicx}	% Including figure files
\usepackage{amsmath}	% Advanced maths commands
\usepackage{soul}
\usepackage{xcolor}

\title[Can we reveal the composition of white dwarfs?]{Can we reveal the core-chemical composition of ultra-massive white dwarfs through their magnetic fields?}

\author[Camisassa et al.]{
Maria E. Camisassa,$^{1}$\thanks{E-mail: maria.camisassa@colorado.edu} 
    Roberto Raddi,$^{2}$
	Leandro G. Althaus,	$^{3,4}$
    Jordi Isern,$^{5,6,7}$
    Alberto Rebassa-Mansergas,$^{2,6}$
\newauthor	Santiago Torres,$^{2,6}$
	Alejandro H. C\'orsico,$^{3,4}$
	and Lydia Korre $^{1}$ 
	\\
$^{1}$ Applied Mathematics Department, University of Colorado, Boulder, CO 80309-0526, USA \\
$^{2}$ Departament de F\'\i sica, 
           Universitat Polit\`ecnica de Catalunya, 
           c/Esteve Terrades 5, 
           08860 Castelldefels, 
           Spain \\
$^{3}$ Instituto de Astrof\'isica de La Plata, UNLP-CONICET, 
           Paseo  del Bosque s/n, 1900 
           La Plata, 
           Argentina \\
$^{4}$  Facultad de Ciencias Astron\'omicas y Geof\'{\i}sicas, Universidad Nacional de La Plata, 
           Paseo del Bosque s/n, 1900 
           La Plata, 
           Argentina\\
$^{5}$ Institute of Space Sciences (ICE, CSIC), Campus UAB, Carrer de Can Magrans s/n, 08193 Barcelona, Spain \\
$^{6}$ Institut d'Estudis Espacials de Catalunya (IEEC), c/Gran Capit\`a 2-4, Edif. Nexus 201, 08034 Barcelona, Spain\\
$^{7}$ Fabra Observatory (RACAB), Rambla dels Estudis 115, 08002 Barcelona, Spain
}

\date{Accepted 2022 July 13. Received 2022 July 12; in original form 2022 March 24}

\pubyear{2015}

\begin{document}
\label{firstpage}
\pagerange{\pageref{firstpage}--\pageref{lastpage}}
\maketitle

\begin{abstract}
Ultra-massive white dwarfs ($ 1.05 \rm M_\odot \lesssim M_{WD}$) are particularly interesting objects that allow us to study extreme astrophysical phenomena such as type Ia supernovae explosions and merger events. Traditionally, ultra-massive white dwarfs are thought to harbour oxygen-neon (ONe) cores. However, recent theoretical studies and new observations suggest that some ultra-massive white dwarfs could harbour carbon-oxygen (CO) cores. Although several studies have attempted to elucidate the core composition of ultra-massive white dwarfs, to date, it has not been possible to distinguish them through their observed properties. Here, we present a new method for revealing the core-chemical composition in ultra-massive white dwarfs that is based on the study of magnetic fields generated by convective mixing induced by the crystallization process. ONe white dwarfs crystallize at higher luminosities than their CO counterparts. Therefore, the study of magnetic ultra-massive white dwarfs in the particular domain where ONe cores have reached the crystallization conditions but CO cores have not, may provide valuable support to their ONe core-chemical composition, since ONe white dwarfs would display signs of magnetic fields and CO would not. We apply our method to eight white dwarfs with magnetic field measurements and we  suggest that these stars are candidate ONe white dwarfs. %We suggest that efforts should be placed in measuring magnetic fields for ultra-massive white dwarfs in this particular domain.
%250 words (200 words for Letters). It has 201
\end{abstract}


%%%%%%%%%%%%%%%%%%%%%%%%%%%%%%%%%%%%%%%%%%%%%%%%%%

%%%%%%%%%%%%%%%%% BODY OF PAPER %%%%%%%%%%%%%%%%%%

\section{Introduction}

White dwarf (WD) stars are the most common end-point of stellar evolution. Therefore, the WD population in our Galaxy is considered a powerful tool to investigate a wide variety of astrophysical problems, from the formation and evolution of our Galaxy to the ultimate fate of planetary systems \citep[see][for a review]{2010A&ARv..18..471A}. 
Among all the WD stars, of special interest are those WDs with masses larger than $\sim 1.05 \rm M_\odot$, commonly referred to as 'ultra-massive' WDs. Observations of these stars have been
largely reported in the literature \citep{2021MNRAS.503.5397K,Hollands2020,2021Natur.595...39C,2022ApJ...926L..24M}, and 
are considered of special interest as they are related to stellar merger episodes, the occurrence of physical processes in the
super asymptotic giant-branch (SAGB) phase, and type Ia
Supernovae and micronovae explosions. 

Despite the large interest in ultra-massive WDs, there is not a clear consensus on the core-chemical composition of such objects. Traditionally, these stars are thought to be born as a result of the isolated evolution of massive intermediate-mass stars with masses larger than $6-9 \, \rm M_\odot$, which reach the SAGB with a degenerate core that develops temperatures high enough to ignite carbon. The violent carbon flame propagates through the core, leading to the formation of an ultra-massive WD with a core composed mainly by oxygen (O) and neon (Ne) \citep{1994ApJ...434..306G,2010A&A...512A..10S}. However, recent theoretical models of the isolated progenitors of ultra-massive WDs suggest that they can avoid carbon burning on the SAGB, leading to the formation of ultra-massive WDs with cores composed mainly by C and O \citep{ALTUMCO2021}. An alternate scenario
for the formation of ultra-massive WDs
has gained relevance in the last years. It is thought that
a relevant fraction of the  single massive WDs
in our Galaxy have been formed as a result of stellar mergers \citep{2020A&A...636A..31T,2020ApJ...891..160C,2022MNRAS.511.5462T}.
Although this scenario has been explored in several hydrodynamical simulations \citep{2007MNRAS.380..933Y,2009A&A...500.1193L}, the core-chemical composition of single ultra-massive WDs born in stellar mergers is a matter of debate. \cite{2021ApJ...906...53S} simulated the merger of two CO WDs and found that that, if the remnant has a mass larger than $\sim 1.05$ M$_\odot$, then the initial CO core is converted into ONe. On the other hand, \cite{2022arXiv220202040W} studied the evolution of the remnant of a merger of a massive CO-core WD with a He-core WD and showed that this scenario can lead to the formation of ultra-massive CO-core WDs with masses up to $\sim 1.20$ M$_\odot$.

 Several studies in the literature have tried to distinguish ONe from CO ultra-massive WDs. \cite{Hollands2020} observed a $1.14 \rm M_\odot$ WD with a unique hydrogen/carbon atmosphere composition. These authors tried to use the atmospheric non-detection of O to test
the core-chemical composition of this WD. However, despite the distinct core-structures, the outer layers of CO- and ONe-core WDs
are almost indistinguishable and, therefore, the degree of O dredge-up does not depend
on the core-composition. A promising avenue to infer the inner chemical stratification of ultra-massive WDs is through Asteroseismology \citep{2019A&ARv..27....7C,2019A&A...632A.119C,
2021A&A...646A..60C}. Nevertheless, in order to pursue such task it is necessary to find a large number of pulsating
ultra-massive WDs with
abundant detected pulsation periods.
 
Despite the persisting efforts, the inference of the core composition of ultra-massive WDs has not been possible to date. Here, we present a new method to accomplish this goal, by using the magnetic field measurements in ultra-massive WDs. According to two-component phase diagrams \citep{2010PhRvL.104w1101H,2020A&A...640L..11B,2010PhRvE..81c6107M} for both CO and ONe mixtures, during the crystallization process, the solid core is enriched in the heavier element (i.e. O in a CO mixture, Ne in an ONe mixture). Therefore, the liquid mantle surrounding the solid core will be depleted in the heavier element, thus having a molecular weight lower than the regions above it and hence inducing a Rayleigh-Taylor mixing in a significant portion of the WD. 
The pioneering study of 
\cite{2017ApJ...836L..28I} suggested that the compositionally driven convection occurring in these stars could drive dynamo magnetic fields similar to those found in Solar System planets. The studies of 
\cite{2021NatAs...5..648S,2021MNRAS.506L..29S} showed that these crystallization-induced dynamos can generate
magnetic fields in cataclysmic variables and in cold metal polluted WDs, explaining their large fraction among the observed population and %\cite{2022arXiv220212902G} showed that crystallization-driven
%dynamos can explain some of the magnetic fields measured for single WDs and WDs in cataclysmic variables and  
\cite{2021MNRAS.505L..74B} demonstrated that such dynamos can account for the observed rareness of bright intermediate polars in globular clusters. Furthermore, 
\cite{2021MNRAS.507.5902B} have analyzed the complete volume-limited WD sample within $20\,$pc from the Sun, showing that the occurrence of magnetic fields is significantly higher in
WDs that have fully or partially crystallized cores than in WDs that have not started the crystallization process yet.
Here, we propose that crystallization-induced dynamo magnetic fields can be used to infer the core-chemical composition of ultra-massive dwarfs. As previously found in \cite{2021A&A...649L...7C,2022MNRAS.511.5198C} and \cite{2020ApJ...902...93B}, ONe-core WDs crystallize at higher luminosities and effective temperatures than their CO-core counterparts, due to the larger coulomb interactions between their ions. Therefore, we expect a particular region in the color-magnitude diagram where ONe ultra-massive WDs have started the crystallization process, and hence could be dominated by 
crystallization-driven dynamos, but CO WDs have not. By analyzing eight magnetic WDs in this particular domain, we propose that the presence of magnetic fields in WDs located within that region is evincing their ONe core-chemical composition.
%\st{We have analyzed eight magnetic WDs in this particular domain.}  %{\bf \st{Although we cannot rule out the possibility that the origin of the magnetic field in these eight WDs could be other than the crystallization-driven dynamo, the relatively good agreement we found suggests that their cores are likely composed by ONe.}}

%This paper is organized as follows. In Section \ref{sec:sample} we describe the sample of selected magnetic WDs that we consider to test our method. In Section \ref{models} and \ref{dynamo} we describe the theoretical models employed to 
%analyze these stars
%and we discuss their soundness. Finally, in Section \ref{Res} we summarize the results and conclude with final remarks.

\section{The white dwarf sample}
\label{sec:sample}

\begin{figure}
	\includegraphics[width=\columnwidth]{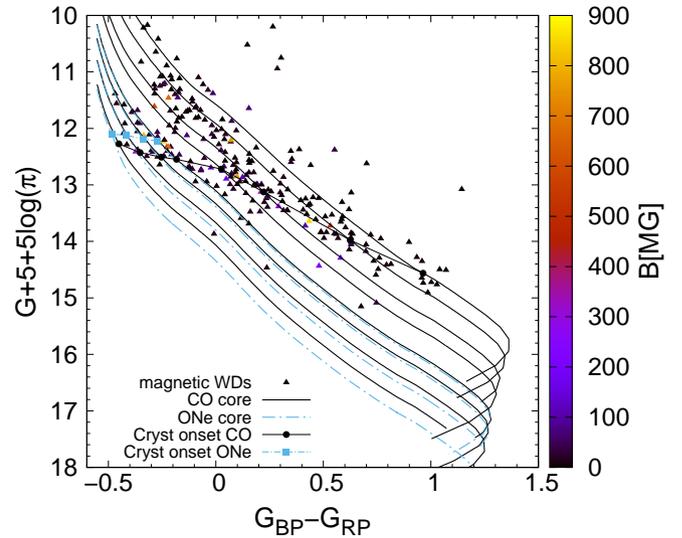}
    \caption{All known magnetic WDs in the {\it Gaia} EDR3 color-magnitude diagram taken from the Montreal White Dwarf Database. Color coding indicates the magnetic field intensity in $10^{6}$ G. The DA CO-core WD evolutionary models with 0.53, 0.66, 0.82, 0.95, 1.10, 1.16, 1.23 and 1.29 M$_\odot$ from \protect \cite{2016ApJ...823..158C,2022MNRAS.511.5198C} are indicated using solid black lines. The DA ONe-core WD evolutionary models with 1.10, 1.16, 1.23 and 1.29 M$_\odot$ from  \protect \cite{camisassa2019} are displayed using dot-dashed blue lines. The black circles and blue squares indicate the crystallization onset in each sequence, respectively.}
    \label{fig:all}
\end{figure}

We selected all known magnetic WDs that are listed in the Montreal White Dwarf Database
 \citep{2017ASPC..509....3D}. These stars are shown in the {\it Gaia} EDR3 color-magnitude diagram in Fig. \ref{fig:all}, together with the hydrogen-rich CO-core and ONe-core WD evolutionary models of \cite{2016ApJ...823..158C,camisassa2019,2022MNRAS.511.5198C}.  We have employed the model atmospheres of \cite{2010MmSAI..81..921K} and \cite{2019A&A...628A.102K} to convert these evolutionary models into the magnitudes in {\it Gaia} EDR3 passbands. Crystallization onset in these models is determined by the phase diagrams of \cite{2010PhRvL.104w1101H} and \cite{2010PhRvE..81c6107M} and is shown as black circles and blue squares, respectively. Colour coding indicates the intensity of the magnetic fields  \citep[see][for reviews on magnetic WDs]{2015SSRv..191..111F,2020AdSpR..66.1025F}. 
The magnetic ultra-massive WDs that have larger luminosities than the ONe crystallization onset are not expected to be crystallizing and, therefore, their magnetic fields must be originated by  mechanisms other than crystallization-driven dynamos.

\begin{figure}
	\includegraphics[width=\columnwidth]{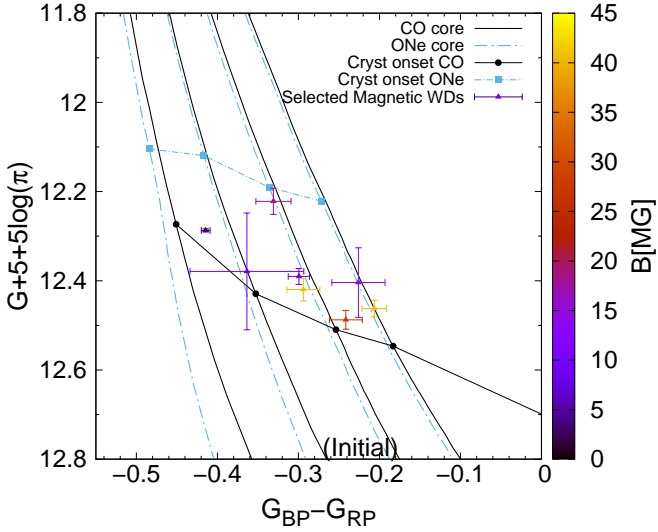}
    \caption{Zoomed-in view of the ONe crystallization region for ultra-massive WDs in the Gaia color-magnitude diagram. 
  That is, we plot those WDs that, if they harbour ONe cores, they have started the crystallization process, but if they harbour CO cores, they have not.
 The adopted colours and symbols are the same as in Fig. \ref{fig:all}. The error bars account for photometric and parallax uncertainties.
  }
    \label{fig:selected}
\end{figure}

Among all the magnetic WDs, we identified eight objects that lie in the region delimited by the CO and ONe crystallization onset and by the 1.29$\,\rm M_\odot$ and 1.10$\,\rm M_\odot$ evolutionary sequences. If these ultra-massive WDs harbour an ONe-core, they are undergoing the crystallization process.  Otherwise, if they have a CO-core, they have not started this process yet.
 This set of eight WDs is shown on the {\it Gaia} color-magnitude diagram in Fig. \ref{fig:selected}. All the selected objects are fully within our region of interest, except for SDSS J112030.34-115051.1 that has larger error bars. 
 All of our selected WDs have hydrogen dominated envelopes and only one of them, WD$1628 + 440$, has an estimation for its rotation period, which has a lower limit of 6hr and an upper limit of 4d \citep{2013ApJ...773...47B}.  The names and properties of these selected WDs are listed in Table \ref{tab}. 

\section{The white dwarf models}
\label{models}
\begin{figure}
	\includegraphics[width=\columnwidth]{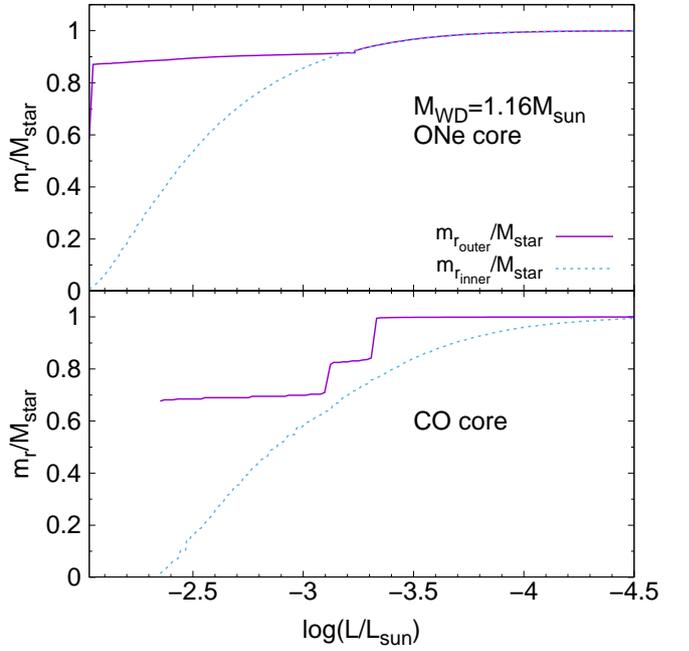}
    \caption{Top panel: Solid purple (dotted blue) line displays the mass fraction at the outer (inner) boundary of the convective region in terms of the logarithm of the stellar luminosity, for our 1.16 M$_{\odot}$ ONe-core WD model. Bottom panel: Same as top panel, but for our 1.16 M$_{\odot}$ CO-core WD model.}
    \label{fig:mass}
\end{figure}

We have employed the ONe-core ultra-massive evolutionary models of \cite{camisassa2019} to analyze the selected magnetic WDs. These models were calculated using the La Plata stellar evolutionary code {\tt LPCODE} \citep[see][for details]{2005A&A...435..631A,2015A&A...576A...9A}. This evolutionary code
has been amply used in the study of different aspects of low-mass star evolution \citep[see][and references therein]{2016ApJ...823..158C,2017ApJ...839...11C,ALTUMCO2021}.
These evolutionary models include a realistic treatment of the crystallization process for ONe mixtures, using an updated phase diagram \citep{2010PhRvE..81c6107M}, and include the energy released as latent heat and by the phase separation process induced by crystallization. 
Furthermore, the initial chemical profiles are the result of the full progenitor evolution through the thermally pulsing SAGB, calculated by \cite{2010A&A...512A..10S}. These accurate initial chemical profiles play a key role in determining the extension of the Rayleigh-Taylor convective region that will be responsible for generating the dynamos. According to the phase diagram employed, in an ONe WD, the solid core is enriched in Ne and, therefore, the immediate surrounding liquid layer is depleted in it, leading to a convective region that takes place in a large portion of the star. The extension of this convective region in the 1.16M$_\odot$ ONe-core model as a function of the logarithm of the stellar luminosity is shown in the top panel of Figure \ref{fig:mass}. The inner (outer) boundary of this region is displayed using a dotted blue (solid purple) line. The inner boundary moves outwards as  crystallization proceeds, whereas the outer boundary barely moves. These boundaries are determined by the chemical abundances that result from the calculation of the full progenitor evolution.
%When $\log(L/L_\odot)\sim -3.25$, more than 90\% of the WD mass has crystallized and the crystallization front reaches the carbon rich layers. Then, we stop modeling the phase separation process, since a three-component phase diagram of C-O-Ne would be needed. Therefore, at this point, the inner and outer boundaries of the mixing region encounter. 

The bottom panel of Figure \ref{fig:mass} shows the extension of the convective region in the 1.16M$_\odot$ CO model of \cite{2022MNRAS.511.5198C} as a function of the logarithm of the stellar luminosity. By comparing the top and bottom panels of this figure, we see that for $-2.35 \lesssim \log(L/L_\odot)\lesssim -2$, the 1.16 M$_\odot$ ONe-core model is developing a convective mantle covering up to $\sim 88\%$ of its mass, that is absent in the 1.16 M$_\odot$ CO model.

We have interpolated the magnitude G and the colour index $\rm G_{BP}-G_{RP}$ in our ONe-core WD sequences to obtain the values of the mass, the luminosity, the cooling age, the effective temperature and the 
logarithm of the surface gravity of the target stars. These quantities are listed in Table \ref{tab}. For those WDs whose masses are not encompassed in \cite{camisassa2019}, we have specifically calculated evolutionary models, by scaling the initial chemical profile of the most similar model from \cite{camisassa2019}. On the basis of these models, we obtained the fraction of crystallized mass ($\rm M_s/M_{WD}$) and the thickness of the convective region ($r_o-r_i$), also listed in Table \ref{tab}, among other parameters.

\section{The white dwarf dynamo}
\label{dynamo}

The magnetic Reynolds number and the magnetic Prandtl number are defined as ${\rm Rm}=ul/\eta$ and $\rm Pm=\nu/\eta$, respectively, where $u$ is a typical velocity, $l$ is a typical length, $\eta$ is the magnetic diffusivity and $\nu$ is the kinematic viscosity. The Ohmic decay time can be estimated as $\tau_\Omega\sim R_{\rm WD}^2/\eta$, where  $R_{\rm WD}$ is the WD radius.
\cite{2017ApJ...836L..28I} estimated that, in a typical massive WD, $\rm Rm \sim 10^{14}-10^{15}$, $\rm Pm \sim 0.58$ and $\tau_\Omega$ is much longer than the WD age. Thus, once the magnetic field has been generated by the  convection-driven dynamo, it will be sustained against 
Ohmic dissipation. In that sense, it is practical to use the dynamo scaling theory of
\cite{2010SSRv..152..565C} to estimate the intensity of the magnetic field in terms
of the properties of the convective region, that takes into account the
balance between the Ohmic dissipation and the energy flux (in the rapidly rotating regime). Following \cite{2010SSRv..152..565C} \citep[see also][]{2009Natur.457..167C}, we obtain:

\begin{equation}
\label{eq:scaling}
\frac{B^2}{8\pi}=cf_\Omega\frac{1}{V} \int_{r_i}^{r_o} \left[\frac{q_c(r)\lambda(r)}{H(r)}\right]^{2/3} \rho(r)^{1/3} 4\pi r^2 dr,
\end{equation}
where the integral is the energy of the convective region ($E$),
$B$ is the magnetic field intensity, $c$ is an adjustable parameter, $f_\Omega$ is the ratio between the Ohmic dissipation and the total dissipation, $V$ is the volume of the convective mantle, $r_i$ and $r_o$ are its inner and outer radius, respectively, $q_c$ is the convected energy flux, $H$ is the scale height  and $\lambda$ is the mixing length.
We adopt $f_\Omega=1$ and $\lambda=H$. The energy density $E/V$ was calculated for each of our selected WDs, and it is listed in Table \ref{tab}.

This scaling law is applicable for WDs in the rapidly rotating regime, whereby the Rossby number, which is the ratio of inertial force to Coriolis force, is much lower than 1, and the equipartition field strength is achieved. We estimate the Rossby number %as  ${\rm Ro}=\frac{v_{\rm conv}}{2 \Omega H}$, where $v_{\rm conv}$ is the characteristic convective velocity, $\Omega$ is the rotation rate and $H$ is the characteristic convective lengthscale. Thus, ${\rm Ro}$ can be estimated 
as 
$\rm Ro\sim \frac{P}{t_{\rm conv}}$, where $P$ is the rotation period and  $t_{\rm conv}$ is the convective turnover timescale. 
We cannot measure the convective turnover timescale in the WD interior, but we can provide an estimation for it.
The density contrast between the floating O enriched material and the surrounding liquid ONe mixture is $\Delta\rho/\rho\sim 10^{-4}$ %\citep{1983A&A...122..212M} 
and, therefore, its upwards acceleration is  $a=g\Delta\rho/\rho\sim 2 \times 10^{5} \rm cm/s^2$. Assuming a limiting velocity of the turbulent eddies as in \cite{2017ApJ...836L..28I},
%$v_{\rm conv}=\sqrt{(3/8)c_baD_b}$, where $D_b$ is the radius of the bubble, considered as $0.1R_{\rm core}$
%\citep{1994GeoJI.117..394M}, and $C_b=2$  in the spherical case \citep{1995ApJ...454..895G}, 
we obtain a characteristic convective velocity of $3\times 10^{5} \rm cm/s$.
Therefore, we estimate the convective turnover timescale for ONe composition to be $\sim 500$ s. 
Measuring the rotation period in WDs is a difficult task. Among the eight WDs analyzed in this paper, only WD 1658+440 has a poorly constrained rotation period with an upper limit of 4 days, which suggests that this star is not in a fast-rotating regime. 
%However, \cite{2022arXiv220212902G} claimed that the density contrast can be lower than the total contrast, yielding longer convective turnover timescales than our prediction.  Therefore, we cannot say whether our selected stars are in the fast rotating regime. 
%, thus implying that the convective turnover timescale is about $\sim 10^{6}-10^{7}$s. Therefore, any WD with a rotation period shorter than $\sim 150$ days would be in the rapidly rotating regime. 

\begin{figure}
	\includegraphics[width=\columnwidth]{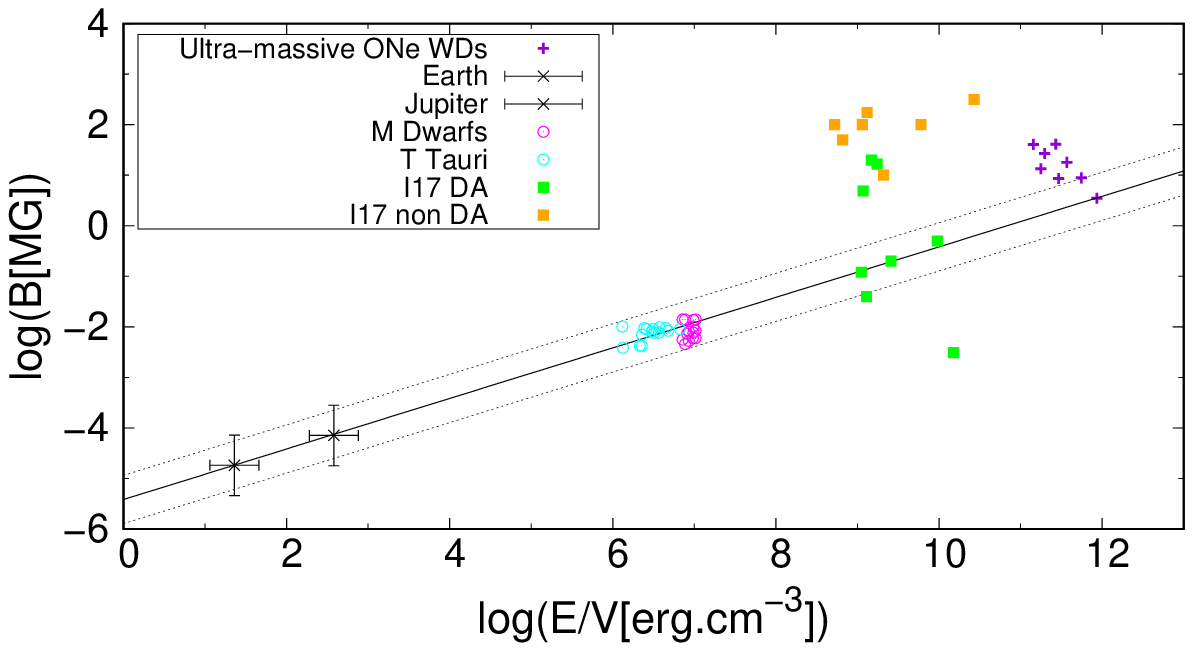}
    \caption{  Magnetic field intensity in terms of the dynamo energy density. Earth and Jupiter are indicated using black crosses with error bars, whereas M dwarfs and T Tauri are plotted using  magenta and cyan circles, respectively \protect \cite{2009Natur.457..167C}. The DA and non-DA WDs from  \protect \cite{2017ApJ...836L..28I} are shown using green and orange squares, respectively.  The ultra-massive WDs  listed in Table \protect \ref{tab} are plotted using purple crosses. The solid line is the scaling law from \protect \cite{2009Natur.457..167C} for the Earth, Jupiter, T Tauri and M dwarf stars considering $c=0.63$, and the dotted lines allow for a deviation of a factor of 3 from it.}
    \label{fig:dynamo}
\end{figure}

 The dynamo energy density and the measured magnetic field intensity of our selected ultra-massive WDs are plotted in Figure \ref{fig:dynamo}, together with the predictions for the Earth, Jupiter, M dwarfs and T Tauri  stars \citep{2009Natur.457..167C} and the WDs studied by \cite{2017ApJ...836L..28I}. The scaling law of \cite{2009Natur.457..167C} considering $c=0.63$ and a deviation of a factor 3 from it are displayed using solid and dotted black lines, respectively. Five of our eight ultra-massive WDs in our sample match the scaling law within a factor of 10, and the other three match it within a factor of 25. In particular, for WD 1658+440 the agreement between the observed and predicted magnetic field is remarkably good. This generally good agreement suggests that these eight WDs could have ONe cores, since CO-core models would not have reached the crystallization onset and, hence, they would not hold crystallization-driven dynamos. However, we wish to remark that although the agreement we found for our eight ultra-massive WDs is better than the one found in \cite{2017ApJ...836L..28I}, the
 observed field distribution of our ONe WD candidates (from $\sim 3$ to $\sim 40$MG) is not significantly different from the field distribution of all isolated magnetic WDs, since $\sim 60$ per cent of them have field strengths in the range 3-40MG  \citep[see Figure 11 in][]{2020AdSpR..66.1025F}.

\begin{table*}
	\centering
	\caption{Name, spectral type and magnetic field intensity of our selected ultra-massive WDs, together with our estimations for the mass, luminosity, cooling times, effective temperature, surface gravity, crystallized mass fraction, and length and dynamo energy density of the convective mantle.}
	\label{tab}
	\begin{tabular}{@{}lccccccccccc@{}} 
		\hline
		Name & Type & B & M & $\log(L/L_\odot)$ & $t_{\rm cool}$&  $T_{\rm eff}$ & $\log{g}$ & $\rm M_s/M_{WD}$ & $r_o-r_i$ & E/V& Ref.\\
		& & [MG] & [$\rm M_\odot$] &  & [Gyrs] &  [k] & [cgs] &  & [$\times10^{8}$\,cm] & [$\times10^{11}$\,erg cm$^{-3}$]& \\		
		\hline
		WD 1658+440 & DA & 3.5 & 1.243 & -1.87 & 0.291 & 28763 & 9.19 & 0.14 & $1.47$ & $8.62$& (a)\\
		SDSS J205233.51-001610.6 & DA & 13.42 & 1.097 & -2.36 & 0.504 & 18168 & 8.83 & 0.19 & $1.77$ & $1.77$& (b)\\
		SDSS J080359.93+122943.9 & DA & 40.7 & 1.093 & -2.43 & 0.570 & 17424 & 8.83 & 0.28 & $1.56$  & $1.43$ & (b)\\
		SDSS J155708.02+041156.4 & DA & 41 & 1.165 & -2.20 & 0.430 & 21403 & 8.98 & 0.20 & $1.63$ & $2.70$& (c)\\
		WD 1131+521 & DA & 8.64 & 1.165 & -2.17 & 0.409 & 21690 & 8.98 & 0.16 & $1.71$ & $2.93$&(b)\\
		SDSS J142703.40+372110.5 & DA & 27.04 & 1.133 & -2.36 & 0.536 & 18885 & 8.91 & 0.30 & $1.40$ & $1.97$&(b) \\
		WD 1327+594 & DQAP & 18 & 1.161 & -2.03 & 0.305 & 23392 & 8.96 & 0.01 & $2.34$ & $3.68$&(d)\\
		SDSS J112030.34-115051.1 & DA & 8.9 & 1.22 & -2.01 & 0.359 & 25383 & 9.11 & 0.23 & $1.41$ & $5.58$&(b) \\
		\hline
		\multicolumn{12}{l}{References: (a) \cite{1992ApJ...394..603S}; (b) \cite{2009A&A...506.1341K}; (c) \cite{2015MNRAS.446.4078K}; (d) \cite{2005AJ....130..734V}}\\
	\end{tabular}
\end{table*}

\section{Summary and conclusions}
\label{Res}

Despite the large interest in ultra-massive WDs, there is not a clear consensus in the literature on whether these stars harbour an ONe or a CO core. Aiming at obtaining information on their core-chemical composition, we have analyzed a sample of eight magnetic ultra-massive WDs selected from the Montreal White Dwarf Database. These stars lie within a particular domain that implies that, if they harbour an ONe core, they are undergoing the crystallization process, whereas if they harbour a CO core, they have not started this process yet. 
That is, ONe core WDs in this region should have a convective region that drives dynamo magnetic fields, but CO WDs should not. 
Relying on precise ONe ultra-massive WD evolutionary models from \cite{camisassa2019} that include realistic chemical profiles that result from the full progenitor evolution, and an updated treatment of the chemical redistribution induced by crystallization, we have modeled these  candidate ONe WDs.
We have estimated their mass, their luminosity, their percentage of crystallized mass, the length of their convective mantles and their convective energy density, among other parameters.

We have compared the measured magnetic fields of these stars with the estimations of the dynamos predicted by the scaling law of \cite{2009Natur.457..167C}, which is suitable for the Earth, Jupiter, M dwarfs and T Tauri stars in the rapidly rotating regime. Among our sample of eight WDs, only WD 1658+440 has a poorly constrained measurement of the rotating period, with an upper limit of 4 days. Therefore, we cannot predict whether these eight WDs are in the rapidly rotating regime.  %, thus implying that this star is likely in the rapidly rotating regime \citep{2022arXiv220212902G}. 
The magnetic field measurements of WD 1658+440, SDSS J205233.51-001610.6, WD 1131+521, WD 1327+594, SDSS J112030.34-115051.1 match the magnetic field intensity predicted by the scaling law  within a factor of 10, indicating that they will likely harbour ONe cores. In particular, the agreement between the predicted and observed magnetic field for WD 1658+440 is remarkably good, evincing its ONe core.
The magnetic field measurements for SDSS J080359.93+122943.9, SDSS J155708.02+041156.4 and SDSS J142703.40+372110.5 exceed the scaling law by a factor of $\sim 20$. The high magnetic fields in these WDs could be indicating that these stars are very fast rotators. Recent numerical simulations suggest that fast rotating convective dynamos with $\rm Ro\ll
1$ can achieve super-equipartition magnetic fields \citep{2016ApJ...829...92A,2019ApJ...876...83A}.
 In that context, \cite{2022MNRAS.514.4111G} claimed that the density contrast driving the
convective zone could be smaller than the one used in this paper and in
\cite{1983A&A...122..212M} and \cite{2017ApJ...836L..28I}, yielding convective turnover
timescales longer than those used here, opening the possibility to these white
dwarf to be fast rotators with strong magnetic fields.
Furthermore, the scaling law that we use has been determined for planets, M dwarfs and T-Tauri stars, where the magnetic Prandtl number is 5-6 orders of magnitudes smaller than in WDs. Considering a likely dependence of the dynamo efficiency on the magnetic Prandtl number, we could expect that the WD dynamo can generate magnetic fields larger than the predictions of the scaling law \citep[see][for a thorough discussion]{2021NatAs...5..648S}, implying that SDSS J080359.93+122943.9, SDSS J155708.02+041156.4 and SDSS J142703.40+372110.5 can also have an ONe core.

Finally, we cannot disregard that the magnetic fields in these WDs could be fossil fields \citep{2004Natur.431..819B}, generated during the merger of
two WDs \citep{2012ApJ...749...25G} or in a common envelope phase, if the binary companion has been destroyed or is in a wide orbit \citep{2008MNRAS.387..897T}. 
However, % detailed population synthesis studies cannot repro-
%duce the number of stars observationally found (Ferrario et al.2015)
none of these scenarios can account for the number of observed magnetic cataclysmic variables, intermediate polar WDs in globular clusters and cold metal polluted WDs \citep{2021NatAs...5..648S, 2021MNRAS.505L..74B,2021MNRAS.506L..29S},
implying that the crystallization-induced dynamos must take place. %
Furthermore, the analysis of the complete volume-limited WD sample within $20\,$pc from the Sun \citep{2021MNRAS.507.5902B} has shown that the
occurrence of magnetic fields is significantly higher in
WDs that have fully or partially crystallized cores than in WDs with fully liquid cores. In this study,  magnetic WDs account for 20\,per cent of the sample and the presence of magnetic fields in WDs with fully liquid cores is $\sim 11$ per cent, whereas this percentage raises to 
$\sim 30$ per cent for WDs that are undergoing or have undergone the crystallization process. This trend is indicating that the crystallization process could be one important cause for developing magnetic fields in WDs. 
Due to the absence of ultra-massive WDs within the 20-pc sample, such a population analysis should be extended to a larger volume in the Solar neighbourhood. However, the spectral characterization of the volume-complete $100\,$pc is still scarce. Within $100\,$pc from the Sun, the frequency of the occurrence of magnetic fields in  ultra-massive WDs with fully liquid cores is $\sim 21$ per cent, whereas this frequency decreases to $3$ per cent for WDs with partially or fully crystallized cores. This drop does not imply that magnetic fields are more frequent in bright young WDs, but it is the likely consequence of various observational biases. Because the main methods for detecting and measuring magnetic fields take advantage of the Zeeman effect that splits the spectral lines and polarizes them, high signal-to-noise ratio as well as high spectral resolution are needed for the analysis of both weak ($< 1$--2\,MG) and very strong ($\gtrsim 100$\,MG) magnetic fields \citep{landstreet2019,2021MNRAS.507.5902B}. Both requirements, combined to the only recent making of the {\it Gaia} 100-pc sample \citep{Jimenez2018}, have worked against the completeness of an intrinsically fainter sample of ultra-massive magnetic WDs. By extrapolating the population analysis of magnetic WDs within 20-pc to the larger volume, we estimate that there should be nearly 2700 magnetic WDs within $100\,$pc, but  just 125 of them are known so far. An exhaustive search for magnetic fields in all ultra-massive WDs within $100\,$pc would be needed to fully confirm our results. %{ \it However, the reasonably good agreement that we found for the crystallization-dynamo predictions is likely revealing that the analyzed ultra-massive WDs harbour ONe cores and are undergoing the crystallization process. }

 Finally, the fact that some WDs that lie in our region of interest have no detected magnetic fields does not imply that these WDs have a CO core, since these WDs could have ONe cores and be crystallizing, but their dynamos are not efficient enough to generate observable magnetic fields. We suggest that, in order to elucidate the core-chemical composition of ultra-massive WDs, efforts should be placed in measuring magnetic fields for large samples of such stars. We expect that close-to-complete large samples of magnetic WDs will be achieved in the next 5--10 years, thanks to the ongoing and currently developing multi-fibre spectroscopic surveys such as DESI \citep{DESI}, SDSS-V \citep{SDSSV}, WEAVE \citep{WEAVE}, and 4MOST \citep{4MOST}. 

\section*{Acknowledgements}

This work  was supported by the NASA 
grants 80NSSC17K0008, 80NSSC20K0193 and 80NSSC21K0455, the postdoctoral fellowship programme Beatriu de Pin\'os %, funded by the Secretary of Universities and Research (Government of Catalonia) and by the Horizon 2020 programme of research and innovation of the European Union under the Maria Sk\l{}odowska-Curie grant 
agreement No 801370, by AGENCIA through the Programa de Modernizaci\'on Tecnol\'ogica BID 1728/OC-AR, by the PIP 112200801-00940 grant from CONICET, by grant G149 from University of La Plata, by the MINECO grant PID2020-117252GB-I00, by the ESP 2017-82674-R, by Grant RYC-2016-20254 funded by MCIN/AEI/10.13039/501100011033 and by ESF Investing in your future.
This work has made use of data from the European Space Agency (ESA) mission {\it Gaia}, %(\url{https://www.cosmos.esa.int/gaia}), 
processed by the {\it Gaia} Data Processing and Analysis Consortium.% (DPAC, \url{https://www.cosmos.esa.int/web/gaia/dpac/consortium}). 

%%%%%%%%%%%%%%%%%%%%%%%%%%%%%%%%%%%%%%%%%%%%%%%%%%
\section*{Data Availability}

Supplementary material will be shared on request to the corresponding author. 



\bibliographystyle{mnras}
\bibliography{example} % if your bibtex file is called example.bib

%\begin{thebibliography}{99}
%\end{thebibliography}

%%%%%%%%%%%%%%%%%%%%%%%%%%%%%%%%%%%%%%%%%%%%%%%%%%

%%%%%%%%%%%%%%%%% APPENDICES %%%%%%%%%%%%%%%%%%%%%

%\appendix



%%%%%%%%%%%%%%%%%%%%%%%%%%%%%%%%%%%%%%%%%%%%%%%%%%

% Don't change these lines
\bsp	% typesetting comment
\label{lastpage}
\end{document}